\documentclass[runningheads]{llncs}

\usepackage{amsmath}
\usepackage{algorithm} 
\usepackage{algpseudocode}
\usepackage{booktabs}
\usepackage{float}
\usepackage{graphicx}
\begin{document}
\title{Modelling Brain Connectivity Networks by Graph Embedding for Dyslexia Diagnosis}
\titlerunning{Graph Embedding for Dyslexia Diagnosis}
%
\author{ Marco A. Formoso\inst{1} \and 
Andrés Ortiz\inst{1,3} \and
Francisco J. Martínez-Murcia\inst{2,3}\and
Nicolás Gallego-Molina\inst{1}\and
Juan L. Luque\inst{4}}
\authorrunning{Marco A. Formoso et. al}
%
\institute{Department of Communications Engineering, University of Málaga (Spain) \and Department of Signal Theory, Telematic and Communications, University of Granada (Spain) \and Andalusian Data Science and Computational Intelligence Institute (DasCI) \and Department of Developmental Psychology, University of Málaga (Spain)
\email{marco.a.formoso@ic.uma.es}}
\maketitle              
\begin{abstract}
Several methods have been developed to extract information from electroencephalograms (EEG). One of them is Phase-Amplitude Coupling (PAC) which is a type of Cross-Frequency Coupling (CFC) method, consisting in measure the synchronization of phase and amplitude for the different EEG bands and electrodes. This provides information regarding brain areas that are synchronously activated, and eventually, a marker of functional connectivity between these areas.
In this work, intra and inter electrode PAC is computed obtaining the relationship among different electrodes used in EEG. The connectivity information is then treated as a graph in which the different nodes are the electrodes and the edges PAC values between them. These structures are embedded to create a feature vector that can be further used to classify multichannel EEG samples. The proposed method has been applied to classified EEG samples acquired using specific auditory stimuli in a task designed for dyslexia disorder diagnosis in seven years old children EEG's. The proposed method provides AUC values up to 0.73 and allows selecting the most discriminant electrodes and EEG bands.

\end{abstract}
\section{Introduction}
EEGs record the electric field fluctuations generated by the neurons as result of their activity \cite{kirschstein2009source}. These recordings provide a non-invasive way to record the brain activity as it reacts to stimuli and can be split in several subsignals of different frequencies called brainwaves. 
There are several brainwaves defined according to their frequency: Delta (0.5-4)Hz, Theta (4-8)Hz, Alpha (8-13)Hz, Beta (13-30)Hz, Gamma (30-100)Hz and Epsilon (100-200)Hz. How and where this bands interact between them in the brain can be measured and used to gain insights on how the brain behaves. We quantified this interaction in terms of phase-amplitude coupling, i.e how the phase of one signal is synchronized with the amplitude of another \cite{canolty2010functional}. 

In this paper we extract the PAC values from the EEGs and are then used to built a graph where every node of this graph correspond to an electrode from the EEG montage. Graphs are powerful mathematical tools that combined with deep learning result in new methods capable of gathering knowledge. Among them is node2vec \cite{grover2016node2vec}, a technique derived from natural language processing (NPL) where the node information and its neighbours is embedded into a N-dimensional vector. We use this embeddings to classify subjects who suffer from developmental dyslexia while performing a selection of the best bands and electrodes.

Developmental Dyslexia (DD) is a learning disorder affecting between 5\% and 13\% of the population \cite{peterson2012developmental}. Early diagnosis of DD in children results essential for their correct intellectual and emotional development. However, it is an challenging task since usually the diagnosis is made after behavioral tests that depends on the child's motivation and could only be applied to reading and writing children since these tests include several tasks based on this skills. The need for diagnosis in pre-reading children has led to the use of biomedical signals without such requirements and disadvantages as EEGs. Thus, the main aim of this work is to provide an effective tool for objective diagnosis using EEG signals. 

The rest of the document is organized as follows. In section 2 we present the data and how it was obtained along with the methods used. Then, in section 3 the results are discussed and finally, in section 4, we present the conclusions and the future work.

\section{Materials and Methods}

\subsection{Database and stimulus}
EEG data used in this work was provided by the Leeduca Reserch Group at the University of Málaga ~\cite{ortiz2020dyslexia}. EEG signals were recorded using the Brainvision acticHamp Plus with 32 active electrodes (actiCAP, Brain Products GmbH, Germany) at a sampling rate of 500 Hz during 15 minute sessions, while presenting an auditory stimulus to the subject. A session consisted of a sequence of white noise stimuli modulated in amplitudes at rates 4.8, 16, and 40 Hz presented sequentially for 5 minutes each.

The database is composed of Forty-eight participants, including 32 skilled readers (17 males) and 16 dyslexic readers (7 males) matched in age (t(1) = -1.4, p $>$ 0.05, age range: 88-100 months). The mean age of the control group was 94, 1 $\pm$  3.3 months, and 95, 6 $\pm$ 2.9 months for the dyslexic group. All participants are
right-handed Spanish native speakers with no hearing impairments and normal or corrected–to–normal vision. Dyslexic children all received a formal diagnosis of dyslexia in the school. None of the skilled readers reported reading or spelling difficulties or have received a previous formal diagnosis of dyslexia. The locations of 32 electrodes used in the experiments is in the 10–20 standardized system, whose names are shown in Figure \ref{fig:pac_matrix}.

\subsection{Signal prepocessing}
Recorded EEG signals were processed to remove artifacts related to eye blinking using the EOG channel and impedance variation due to movements. This was addressed using blind source separation by means of Independent Component Analysis (ICA). Then, EEG signal of each channel was referenced to the Cz electrode and normalized independently to zero mean and unit  variance. Baseline correction was also applied. 

\subsection{Phase-Amplitude Coupling (PAC)}
Phase-Amplitude Coupling is a type of Cross-Frequency Coupling. In Canolty et al. \cite{canolty2010functional} they propose that CFC arises as a sort of cross-domain connectivity, where the high-frequency oscillations reflect local domain cortical processing and the low-frequency rhythms are \textit{dynamically entrained across distributed brain regions}. Several measures have been proposed over time to quantify this connectivity and compared in H{\"u}lsemann et al. \cite{hulsemann2019quantification} (Phase-Locking Value, Mean Vector Length (MVL), Modulation Index (MI), and Generalized-Linear-Modeling-Cross-Frequency-Coupling). Their recommendation is to use MVL-based descriptors for signals recorded at high sample rate and high signal-noise ratio. In addition they advise to use MI as it complements the weaknesses of MVL. The main issue with MVL is its dependency on the amplitude of the signal providing amplitude so amplitude outliers can strongly affect the final result of the measure although these issues can be partially addressed by means of permutation testing. 

PAC is a way to quantify the interaction between the phase of low-frequency components and the amplitude of high-frequency components in a EEG. To this end, the first step consist in decomposing the signal into its frequency components. The classical approach using Fourier methods require the signal to meet the requirements of periodicity and stationarity, both of which are not present in EEG signals. Instead, Hilbert transform is used posterior to a band-pass filter for every brainwave frequency range, obtaining the corresponding analytic signal to derive the amplitude, phase and frequency of the signal over time. As explained in the following, phase-synchronization measurements can be computed from the instantaneous, unwrapped phase. 

As previously stated, there are several methods to quantify the synchronization of two signals. The selected one in this work is MVL introduced in \cite{canolty2006high} and defined as follows:

\begin{equation}
\label{eq:MVL}
MVL = \left| \frac{\sum_{t=1}^{n}a_t e^{i(\theta_t)}}{n} \right|
\end{equation}

where $n$ is the total number of data points, $a_t$ is the amplitude at point $t$ of the signal providing amplitude and $\theta_t$ is the phase angle at point $t$ of the signal providing phase. Additionally, spurious coupling values are discarded by a permutation test using surrogates computed by swapping amplitude values in the former signal \cite{aru2015untangling}.

\subsection{Connectivity Estimation by Phase-Amplitude Coupling}
\begin{figure}[!htb]
    \centering
    \includegraphics[scale=0.6]{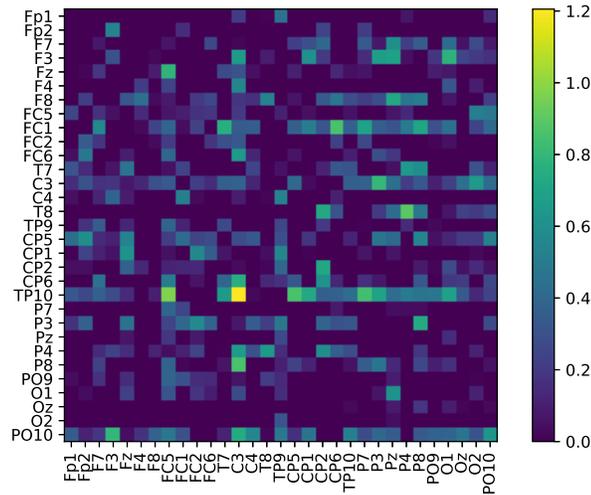}
    \caption{PAC Matrix for Delta-Theta. The nodes on the Y axis are the nodes for Delta band providing phase and the nodes on X axis are the nodes for Theta band providing amplitude. Higher values represent a higher synchronization between bands. }
    \label{fig:pac_matrix}
\end{figure}

Once MVL values are computed, a connectivity matrix can be defined as shown in Figure \ref{fig:pac_matrix}. This matrix shows inter and intra node connectivity allowing the composition of a directed graph by using the compute PAC matrix as an adjacency matrix. The values represent the degree of synchronization between bands and the weight of the edges when the graph is built.

\subsection{Graph Embedding}
Several algorithms have been recently developed in order to extract information from graphs. They can be classified into different categories depending on the technique to process the connectivity matrix that represents the graph: factorization, random walk and deep learning \cite{goyal2017graph}. \textit{node2vec} is in the group of random walks although it uses a NPL embedding method we commonly encounter in deep learning architectures called word2vec \cite{mikolov2013efficient}. In word2vec as in node2vec we try to find a representation of the word/node while preserving as much information as possible. As a result, similar objects have similar embeddings. 
In this context, and to show the equivalence to NPL, the graph can be seen as a document of which we can extract sentences. These sentences are composed by a finite set of nodes.

The selection of nodes in order to build a sentence in node2vec is performed by means of a random walk through the graph: starting in a node, we select one of its neighbour with bias (\ref{eq:walk_prob}) and it is added to the set. The same operation is repeated but starting from the last added node. The number of repetitions of this operation is called walk length and it is one of the hyperparameters of node2vec algorithm. In addition, there are two more hyperparameters, $q$ and $p$ that control how the graph is traversed. If the walk is in a node $v$ coming from a node $x$ and $t\in neighbours(v)$, being $d_{tx}$ the distance from $x$ to $t$, the likelihood $\alpha_{pq}(t,x)$ of adding $t$ to the set is as follows:

\begin{equation}
\label{eq:walk_prob}
\alpha_{pq}(t,x) 
\begin{cases} 
\frac{1}{p} & \text{if }  d_{tx} =0 \\
1           & \text{if }  d_{tx} =1 \\
\frac{1}{q} & \text{if }  d_{tx} =2 \\
\end{cases} 
\end{equation}

This procedure is applied for all of the neighbours $t$ of $v$. The number of walks is another adjustable parameter and sets the number of walks which will be available to train the aforementioned skip-gram model thus obtaining the embedding for each node.

\subsection{Classification}
\begin{figure}[!htb]
    \begin{center}
    \includegraphics[scale=0.6]{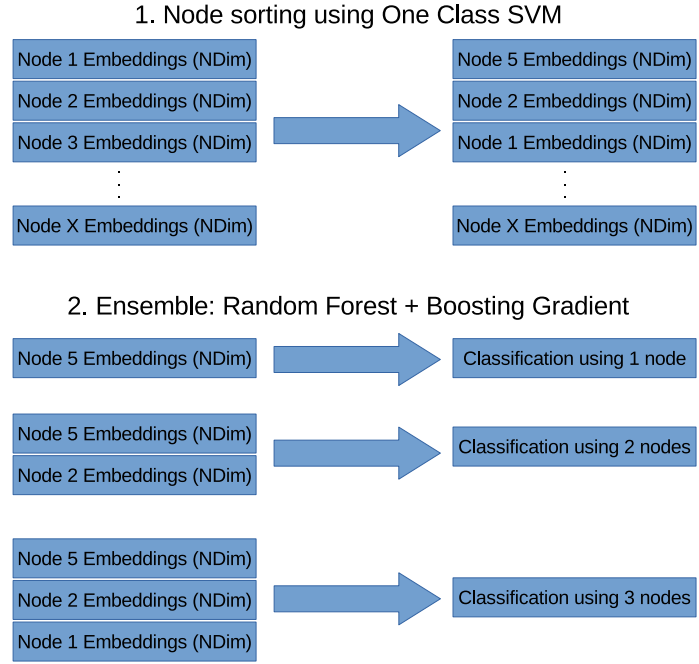}
    \caption{Classification. This process is done for every embedding dimension N ranging from 1 to 9.}
    \label{fig:class_steps}
    \end{center}
\end{figure}

As a first step, node selection is performed by a one-class support vector machine (OCSVM) which is typically used to detect outliers in a dataset. In this work, for the embeddings of every single node, a OCSVM is trained with control samples and it is then used to predict dyslexia samples. In this way, we can determine the most discriminant nodes (i.e. those which detect a larger number of outliers). Then, they are sorted based on the numbers of detected outliers, and a classification task is performed using an ensemble classifier composed of a random forest and a gradient boosting. This can be seen in Algorithm \ref{alg:classification} as well as in Fig. \ref{fig:class_steps}.

\begin{algorithm}[!htb]
	\caption{Classification} 
	\label{alg:classification}
	\begin{algorithmic}[1]
	\State $Nodes =$ Set of nodes
	\State $Embeddings$ = Data with shape ( \#Subjects ,\#Nodes, EmbeddingDimension) 
	\State $Outliers = [ ]$
	
	\For {$node$ \textbf{in }$Nodes$}
	    \State  $Outliers.append( GetNumberOutliers( node ) )$
	\EndFor
	\State $Indices = Outliers.argumentsort()$
	\For { $index$ \textbf{from} $0$  \textbf{to} $Length(Indices)$}
	   \State $DataSet = Embeddings[:, Indices[:index]]$
	   \State $VotingClassifier(Dataset)$
	\EndFor	
	\end{algorithmic} 
\end{algorithm}

\section{Results}

In this section, experimental results through the entire classification pipeline are shown, using PAC to compose the connectivity matrix and embedding the corresponding graph.
PAC is obtained with the tensorpac library\cite{combrisson_tensorpac_2020}. As explained in the introduction, PAC is used to measure the synchronization between the phase of one band and the amplitude of another. Typically, PAC is computed between a low frequency band providing phase and a higher frequency band providing amplitude meaning it is not necessary to calculate all the permutations for the bands. We computed the PAC between the following bands:
\begin{itemize}
    \item Bands providing phase: Delta, Theta, Alpha, Beta
    \item Bands providing amplitude: Beta, Gamma, Epsilon
\end{itemize}

As a result, we obtain twelve $31x31$ PAC matrices for each subject like the one in Figure \ref{fig:pac_matrix}. These matrices are used as adjacency matrices to construct a directed graph that is then processed with the node2vec algorithm. Moreover, experiments varying the node embedding dimension from 1 to 9 have been carried out, as it is related to the representation capability of the embedding. In the classification step, the hyperparameters max\_depth=3 and n\_estimators=20 for the boosting gradient classifier were selected while the random forest hyperparameters were set to n\_estimators=100.

\begin{table}[!htb]
\centering
\caption{Maximum AUC obtained for embeddings from 1 to 9 dimensions}
\label{table:auc}
\setlength{\tabcolsep}{6pt}
\begin{tabular}{lrrrrrrrrr}
\toprule
{} &    1 &    2 &    3 &    4 &    5 &    6 &    7 &    8 &    9 \\
\midrule
Delta-Beta    & 0.69 & 0.47 & 0.57 & 0.66 & 0.54 & 0.55 & 0.53 & 0.65 & 0.62 \\
Delta-Gamma   & 0.68 & 0.55 & 0.59 & 0.58 & 0.54 & 0.55 & 0.55 & 0.59 & \textbf{0.73} \\
Delta-Epsilon & 0.47 & 0.57 & 0.59 & 0.66 & 0.62 & 0.55 & 0.69 & 0.66 & 0.58 \\
Theta-Beta    & 0.56 & 0.62 & 0.45 & 0.60 & 0.56 & 0.52 & 0.44 & 0.50 & 0.48 \\
Theta-Gamma   & 0.60 & 0.66 & 0.57 & 0.48 & 0.51 & 0.59 & 0.58 & 0.65 & 0.55 \\
Theta-Epsilon & 0.53 & 0.63 & 0.53 & 0.67 & 0.55 & 0.59 & 0.61 & 0.56 & 0.52 \\
Alpha-Beta    & 0.68 & 0.54 & 0.50 & 0.48 & 0.48 & 0.58 & 0.55 & 0.62 & 0.72 \\
Alpha-Gamma   & \textbf{0.72} & 0.62 & 0.65 & 0.53 & 0.57 & 0.44 & 0.61 & 0.55 & 0.62 \\
Alpha-Epsilon & 0.61 & 0.59 & 0.52 & 0.59 & 0.64 & 0.65 & 0.51 & 0.57 & 0.51 \\
Beta-Beta     & 0.58 & 0.55 & 0.66 & 0.60 & 0.46 & 0.55 & 0.58 & 0.56 & 0.53 \\
Beta-Gamma    & 0.59 & 0.64 & 0.59 & 0.68 & 0.67 & 0.59 & 0.64 & 0.67 & 0.60 \\
Beta-Epsilon  & 0.58 & 0.67 & 0.51 & 0.61 & 0.52 & 0.63 & 0.58 & 0.59 & 0.56 \\
\bottomrule
\end{tabular}
\end{table}

\begin{figure}[!htb]
    \begin{center}
    \includegraphics[width=0.70\textwidth]{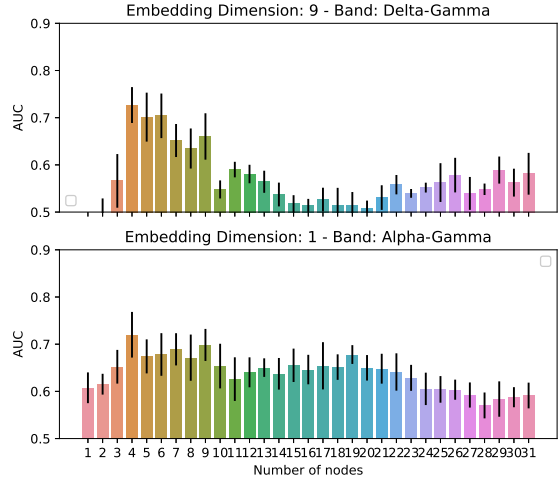}
    \caption{AUC values for Delta-Gamma and Alpha-Gamma}
    \label{fig:nodes_comp}
    \end{center}    
\end{figure}

Maximum AUC values for different embedding dimension and band combinations are shown in Table \ref{table:auc} . The best values are found for 1 and 9  embedding dimensions, with a maximum AUC of 0.73 for the 9 dimension embedding in the Delta-Gamma band. In Figure \ref{fig:nodes_comp} a comparative between these two bands is shown for every number of nodes. The best values are obtained using 4 nodes in both situations. Moreover, it can be seen that not all nodes carry discriminant information, so adding more may be counterproductive. 

\begin{figure}[!htb]
    \begin{center}
    \includegraphics[width=\textwidth]{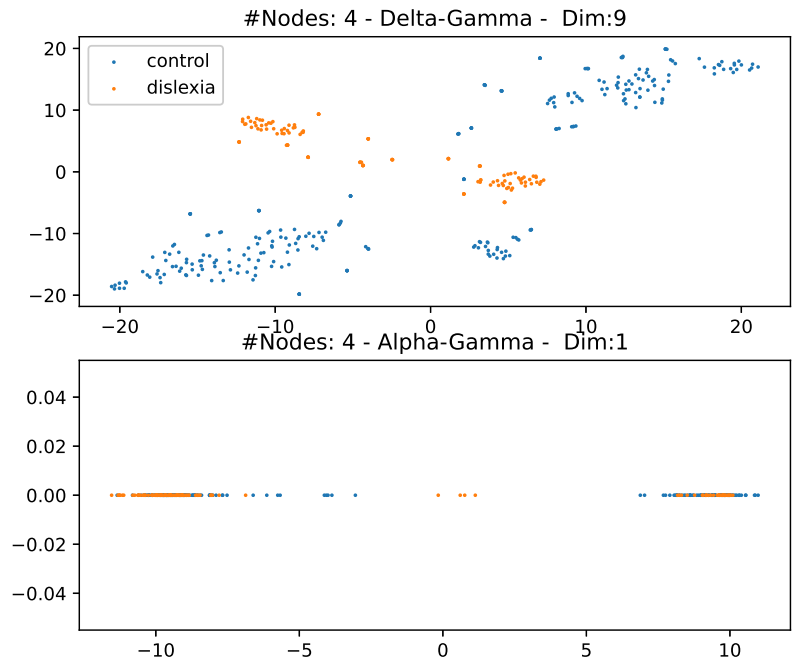}
    \caption{Distribution for Delta-Gamma and Alpha-Gamma}
    \label{fig:scatter_comp}
    \end{center}
\end{figure}

In Figure \ref{fig:scatter_comp} we can see the distribution of the embeddings for Delta-Gamma and Alpha-Gamma. For Delta-Gamma we use TSNE to reduce the 9D embeddings to 2D. We can see clusters for both groups that explain the results. 

Finally, although Delta-Gamma provides the best AUC it is worth noticing that we are dealing with 9 dimension embeddings while getting similar results using only 1D. This is a tradeoff that we will have to take into account for future experiments. 

\section{Conclusions and Future Work}
In this work we present a classification method that relies on PAC-based connectivity to construct directed graph explaining the connectivity patterns found in EEG signals. Moreover, a graph embedding technique was used to compress all the information on the graph in a feature set representing it. Thus, a feature space consisting of descriptors for the graphs of each subject is composed and then used for subject classification. The methodology used demonstrated that graphs computed from PAC matrices provide discriminative enough information to separate dyslexia and control groups, especially in certain EEG bands as Alpha-Gamma and Delta-Gamma with AUC in the 0.7-0.73 range. 

As future work we plan to use another metrics to describe the graph structure and to extract features as well as to use specific deep learning methods for graphs such as graph neural networks.

%
%
%
\bibliographystyle{splncs04}
\bibliography{bibliography}

\begin{thebibliography}{10}
\providecommand{\url}[1]{\texttt{#1}}
\providecommand{\urlprefix}{URL }
\providecommand{\doi}[1]{https://doi.org/#1}

\bibitem{aru2015untangling}
Aru, J., Aru, J., Priesemann, V., Wibral, M., Lana, L., Pipa, G., Singer, W.,
  Vicente, R.: Untangling cross-frequency coupling in neuroscience. Current
  opinion in neurobiology  \textbf{31},  51--61 (2015)

\bibitem{canolty2006high}
Canolty, R.T., Edwards, E., Dalal, S.S., Soltani, M., Nagarajan, S.S., Kirsch,
  H.E., Berger, M.S., Barbaro, N.M., Knight, R.T.: High gamma power is
  phase-locked to theta oscillations in human neocortex. science
  \textbf{313}(5793),  1626--1628 (2006)

\bibitem{canolty2010functional}
Canolty, R.T., Knight, R.T.: The functional role of cross-frequency coupling.
  Trends in cognitive sciences  \textbf{14}(11),  506--515 (2010)

\bibitem{combrisson_tensorpac_2020}
Combrisson, E., Nest, T., Brovelli, A., Ince, R.A.A., Soto, J.L.P., Guillot,
  A., Jerbi, K.: Tensorpac: {An} open-source {Python} toolbox for tensor-based
  phase-amplitude coupling measurement in electrophysiological brain signals.
  PLoS computational biology  \textbf{16}(10),  e1008302 (Oct 2020).
  \doi{10.1371/journal.pcbi.1008302}

\bibitem{goyal2017graph}
Goyal, P., Ferrara, E.: Graph embedding techniques, applications, and
  performance: A survey. Knowledge-Based Systems  (2018).
  \doi{https://doi.org/10.1016/j.knosys.2018.03.022},
  \url{http://www.sciencedirect.com/science/article/pii/S0950705118301540}

\bibitem{grover2016node2vec}
Grover, A., Leskovec, J.: node2vec: Scalable feature learning for networks. In:
  Proceedings of the 22nd ACM SIGKDD international conference on Knowledge
  discovery and data mining. pp. 855--864 (2016)

\bibitem{hulsemann2019quantification}
H{\"u}lsemann, M.J., Naumann, E., Rasch, B.: Quantification of phase-amplitude
  coupling in neuronal oscillations: comparison of phase-locking value, mean
  vector length, modulation index, and
  generalized-linear-modeling-cross-frequency-coupling. Frontiers in
  neuroscience  \textbf{13}, ~573 (2019)

\bibitem{kirschstein2009source}
Kirschstein, T., K{\"o}hling, R.: What is the source of the eeg? Clinical EEG
  and neuroscience  \textbf{40}(3),  146--149 (2009)

\bibitem{mikolov2013efficient}
Mikolov, T., Chen, K., Corrado, G., Dean, J.: Efficient estimation of word
  representations in vector space. arXiv preprint arXiv:1301.3781  (2013)

\bibitem{ortiz2020dyslexia}
Ortiz, A., Martinez-Murcia, F.J., Luque, J.L., Gim{\'e}nez, A., Morales-Ortega,
  R., Ortega, J.: Dyslexia diagnosis by eeg temporal and spectral descriptors:
  An anomaly detection approach. International Journal of Neural Systems pp.
  2050029--2050029 (2020)

\bibitem{peterson2012developmental}
Peterson, R.L., Pennington, B.F.: Developmental dyslexia. The lancet
  \textbf{379}(9830),  1997--2007 (2012)

\end{thebibliography}
\end{document}